\documentclass[useAMS,usenatbib]{mn2e} \usepackage{amsmath,amssymb}
\usepackage{graphicx,mathptm} \usepackage{times} \usepackage{natbib}
\usepackage{psfig}
\usepackage{color}
\newcommand{\Mpch}{$h^{-1}\,\mbox{Mpc}$\,}
\newcommand{\Gpch}{$h^{-1}\,\mbox{Gpc}$\,}

\newcommand{\eg}{{e.g.}}

\title[The BAO signal in cDE models] {Exploiting the shift of baryonic
  acoustic oscillations as a dynamical probe for dark interactions}

\author[Vera Cervantes et al.] {Victor D. Vera
  Cervantes$^{1}$\thanks{E-mail: victor.veracervantes@unibo.it},
  Federico Marulli$^{1,2,3}$, Lauro Moscardini$^{1,2,3}$, \and Marco
  Baldi$^{1}$ and Andrea Cimatti$^{1,2}$\\ \\ $^1$Dipartimento di
  Fisica e Astronomia, Alma Mater Studiorum - Universit\`a di Bologna,
  viale Berti Pichat 6/2, I-40127 Bologna,
  Italy\\ $^2$INAF-Osservatorio Astronomico di Bologna, Via Ranzani 1,
  40127, Bologna, Italy\\ $^3$INFN/National Institute for Nuclear
  Physics, Sezione di Bologna, viale Berti Pichat 6/2, I-40127
  Bologna, Italy\\
}

\begin{document}

\maketitle

\begin{abstract}
The baryonic acoustic peak in the correlation function of galaxies and
galaxy clusters provides a standard ruler to probe the space-time
geometry of the Universe, jointly constraining the angular diameter
distance and the Hubble expansion rate. Moreover, non-linear effects
can systematically shift the peak position, giving us the opportunity
to exploit this clustering feature also as a dynamical probe. We
investigate the possibility of detecting interactions in the dark
sector through an accurate determination of the baryonic acoustic
scale.  Making use of the public halo catalogues extracted from the
{\small CoDECS} simulations -- the largest suite of N-body simulations
of interacting dark energy models to date -- we determine the position
of the baryonic scale fitting a band-filtered correlation function,
specifically designed to amplify the signal at the sound horizon. We
analyze the shifts due to non-linear dynamics, redshift-space
distortions and Gaussian redshift errors, in the range $0\leq z
\leq2$. Since the coupling between dark energy and dark matter affects
in a particular way the clustering properties of haloes and,
specifically, the amplitude and location of the baryonic acoustic
oscillations, the cosmic evolution of the baryonic peak position might
provide a direct way to discriminate interacting dark energy models
from the standard $\Lambda$CDM framework. To maximize the efficiency
of the baryonic peak as a dynamic probe, the correlation function has
to be measured in redshift-space, where the baryonic acoustic shift
due to non-linearities is amplified. The typical redshift errors of
spectroscopic galaxy surveys do not significantly impact these
results.
 
\end{abstract}


%%%%%%%%%%%%%%%%%%%%%%%%%%%%%%%%%%%%%%%%%%%%%%%%%%%%%%%%%%%%%%%%%%%%%%%%%%%%%%%
%%%%%%%%%%%%%%%%%%%%%%%%%%%%%%%%%%%%%%%%%%%%%%%%%%%%%%%%%%%%%%%%%%%%%%%%%%%%%%%

\section {Introduction}
\label{intro}

Understanding the physical origin of the accelerated expansion of the
Universe is one of the main drivers of the current cosmological
research activity.  The so-called $\Lambda$-Cold Dark Matter
($\Lambda$CDM) model is to date the most reliable cosmological
scenario to explain such acceleration, due to its success in matching
the observed properties of the large scale structure (LSS) of the
Universe. However, there are still some observational tensions at
small and intermediate scales, like the lack of luminous satellites in
cold dark matter (CDM) haloes \citep[see
  \eg][]{navarro1996,BoylanKolchin_Bullock_Kaplinghat_2011}, the
observed low baryon fraction in galaxy clusters \citep[see
  \eg][]{ettori2003, mccarthy2007} and the high
velocities detected in the large-scale bulk motion of galaxies
\citep[see \eg][]{watkins2009} or in systems of colliding galaxy
clusters, as the famous ``Bullet Cluster" \citep[see
  \eg][]{lee2010}. It is still unclear whether the above apparent
disagreements are a consequence of some not well understood
astrophysical phenomena at small scales, or if they really represent a
signature of an underlying cosmological scenario different from the
standard $\Lambda$CDM model. In any case, this provides an obvious
motivation to explore alternative cosmological models that could
explain these astrophysical anomalies.  Furthermore, the nature of the
dark sector of the Universe, made of dark energy (DE) -- the component
responsible for the late time acceleration -- and dark matter (DM) --
a kind of matter that negligibly interacts with radiation and can only
be detected by its gravitational influence, is still unknown, even if
it constitutes more than 95\% of the total energy of the Universe.

In the last fifteen years, after the detection of the accelerated 
expansion of the Universe \citep[][]{Riess_etal_1998,Perlmutter_etal_1999,Schmidt_etal_1998}, widespread theoretical and observational efforts have been made 
to clarify whether the observed DE component
is just a cosmological constant $\Lambda$, or if its origin comes
from other sources that can dynamically change in time.  In this work,
we investigate the LSS of the Universe in the so-called coupled (or
interacting) DE models (cDE hereafter). This kind of cosmological scenarios are
based on the dynamical evolution of a classical scalar field, $\phi$,
that plays the role of the DE and interacts with the CDM particles by
exchanging energy-momentum \citep[see
  e.g.][]{wetterich1995, amendola2000, Amendola_2004}. Interestingly, cDE
models can alleviate some observational tensions at small scales,
that are presently not well understood in the $\Lambda$CDM cosmology, as
mentioned above \citep[see
  e.g.][]{baldi2010,baldi_pettorino2011,lee2011}.

Baryonic Acoustic Oscillations (BAO) are the relic imprints left over
by the primordial density perturbations in the early Universe, when
the photon pressure was counteracting the gravitational collapse of
baryons, generating acoustic waves in the baryon-photon plasma. When
the Universe became cool enough to allow the electrons and protons to
recombine, at $z\sim1000$, baryons and photons decoupled and the
acoustic waves became frozen in the baryon distribution. Baryons then
progressively fell into CDM potential wells and CDM was, in turn, also
attracted to baryon overdensities \citep[see \eg][]{peebles1970}. The
BAO feature, corresponding to the scale of the comoving ``sound
horizon", can be detected in the matter distribution as a small excess
in the number of galaxy pairs at $\sim$100 \Mpch, that creates either
a single peak in the correlation function, $\xi(r)$, in configuration
space, or a series of peaks in the power spectrum, $P(k)$, in Fourier
space. As the sound horizon scale is well constrained by the cosmic
microwave background (CMB) observations, the BAO location at different
redshifts provides a standard ruler to measure the expansion history
of the Universe and to constrain the DE equation of state, $w(z)$
\citep[see \eg][]{seo2003, albrecht2006, amendola2012}. The BAO
provide a powerful probe to investigate the nature of the late time
acceleration, as they depend only slightly on the still uncertain
astrophysical phenomena that could introduce systematic bias in the
cosmological constraints.

Several independent measurements of the BAO feature in galaxy redshift
surveys contributed to the great success of the concordance
$\Lambda$CDM model \citep[see \eg][]{eisenstein2005, cole2005,
  hutsi2006, percival2007, padmanabhan2007, okumura2008, sanchez2009,
  gaztanaga2009, percival2010, kazin2010, blake2011, beutler2011}, and
a number of future experiments are planned to use these features for
precision cosmology, such as Euclid\footnote{http://www.euclid-ec.org}
\citep{laureijs2011,amendola2012},
BigBOSS\footnote{http://bigboss.lbl.gov} \citep{schlegel2011}, and the
Wide-Field Infrared Survey Telescope
(WFIRST)\footnote{http://wfirst.gsfc.nasa.gov}.

In order to exploit the BAO as a robust standard ruler, all the
systematic effects have to be kept under control and the position of
the BAO peak has to be measured with high precision. The traditional
estimators, $\xi(s)$ and $P(k)$, have some disadvantages that can
affect the measure of the BAO at different redshifts. The former can
be affected by the integral constraint at large-scale, if the cosmic
number density of the population of mass tracers is not accurately
known. The latter is instead affected by the uncertainty in the
small-scale power, that is difficult to model due to non-linear
effects. To reduce these possible systematics, \citet{xu2010} proposed
a new robust estimator to analyze LSS and BAO, band-filtering the
information in the two-point correlation function. The above estimator
has been recently used to extract cosmological constraints from the
BAO feature in the WiggleZ data \citep{blake2011}.

Non-linearities also affect the BAO as a standard ruler. According to linear
perturbation theory, the sound horizon scale imprinted in the early
Universe remains unaltered as a function of time. However,
non-linearities cause a slight shift of the peak position relative to
the linear prediction \citep[see \eg][]{seo2005, crocce2008,
  angulo2008, padmanabhan2009}, and redshift-space distortions (RSD)
further exacerbate this effect \citep[see \eg][]{smith2008}. As a
consequence, to calibrate the BAO as an unbiased standard ruler, the
non-linear effects have to be accurately quantified in real- and
redshift-space. Moreover, as we will discuss extensively in the next
sections, the way that non-linearities and RSD affect the BAO feature
depends on the assumed cosmological model. As a consequence, the BAO
peak provides also a dynamical probe that could be of some help in
discriminating between the $\Lambda$CDM model and alternative DE
scenarios characterized by a different non-linear regime of structure
evolution.

In this paper, we will focus our investigation on configuration space
and extend to larger scales the work presented in \citet{marulli2012},
analyzing the BAO feature in cDE models with the band-filtered
correlation function of \citet{xu2010}. In particular, we focus on the
shift of the BAO position due to non-linearities and RSD, in the
redshift range $0\leq z \leq 2$. Moreover, we quantify the impact of
Gaussian redshift errors on the BAO feature. To properly describe all
the linear and non-linear effects at work, we employ a series of
state-of-the-art N-body simulations for a wide range of different cDE
scenarios -- the COupled Dark Energy Cosmological Simulation project
({\small CoDECS}) \citep{CoDECS} -- that are used to investigate the
impact that DE interactions can have on LSS, and in particular on BAO,
with respect to the $\Lambda$CDM case.

The structure of the paper is as follows. In \S \ref{cDE} and \S
\ref{N-body}, we describe the cDE models and the exploited set of
N-body experiments analyzed in this work to simulate the LSS in these
cosmologies. In \S \ref{new_stat_th}, we review our
theoretical tools, describing the new statistic, $\omega_0(r_s)$, used
in our investigation. Our results are described in \S \ref{results},
where we analyze the CDM halo clustering at large scales, focusing on
the shift of the BAO signal position through cosmic time. Finally, in
\S \ref{concl} we draw our conclusions.

%%%%%%%%%%%%%%%%%%%%%%%%%%%%%%%%%%%%%%%%%%%%%%%%%%%%%%%%%%%%%%%%%%%%%%%%%%%%%%%
%%%%%%%%%%%%%%%%%%%%%%%%%%%%%%%%%%%%%%%%%%%%%%%%%%%%%%%%%%%%%%%%%%%%%%%%%%%%%%%

\begin{table*}
\begin{center}
\caption{List of cosmological models considered in the {\small CoDECS}
  project and their specific parameters (see text for details).}
\begin{tabular}{llccccc}
\hline \hline Model & Potential & $\alpha$ & $\eta_{0}$ & $\eta_{1}$ &
$w_{\phi}(z=0)$ & $\sigma _{8}(z=0)$ \\ \hline $\Lambda $CDM & $V(\phi
) = A$ & -- & -- & -- & $-1.0$ & $0.809$ \\ EXP001 & $V(\phi ) =
Ae^{-\alpha \phi }$ & 0.08 & 0.05 & 0 & $-0.997$ & $0.825$ \\ EXP002 &
$V(\phi ) = Ae^{-\alpha \phi }$ & 0.08 & 0.1 & 0 & $-0.995$ & $0.875$
\\ EXP003 & $V(\phi ) = Ae^{-\alpha \phi }$ & 0.08 & 0.15 & 0 &
$-0.992$ & $0.967$ \\ EXP008e3 & $V(\phi ) = Ae^{-\alpha \phi }$ &
0.08 & 0.4 & 3 & $-0.982$ & $0.895$ \\ SUGRA003 & $V(\phi ) =
A\phi^{-\alpha }e^{\phi ^{2}/2}$ & 2.15 & -0.15 & 0 & $-0.901$ &
$0.806$ \\ \hline \hline
\end{tabular}
\label{tab:models}
\end{center}
\end{table*}

\section {The Coupled Dark Energy models}
\label{cDE}

The cosmological scenarios investigated in the present work belong to
the class of inhomogeneous DE models with species-dependent couplings
\citep[][]{Damour_Gibbons_Gundlach_1990}.  They have been indroduced
as a possible alternative to the standard $\Lambda$CDM cosmology to
explain the accelerated expansion of the Universe and to alleviate the
``fine-tuning" problems that affect the scenario with a cosmological
constant \citep{wetterich1995, amendola2000}. These models are based
on the dynamical evolution of a DE scalar field, $\phi$, that
interacts with the other components of the Universe by exchanging
energy-momentum during cosmic evolution. In particular, the case of a
coupling between DE and massive neutrinos has been proposed by
\citet{Amendola_Baldi_Wetterich_2008}, while an interaction between DE
and CDM particles has been considered by e.g. \citet{wetterich1995,
  amendola2000, Amendola_2004, Farrar2004,
  Brookfield_VanDeBruck_Hall_2008, Koyama_etal_2009,
  CalderaCabral_2009, baldi2011d, Baldi_2012a}.  Such DE-CDM
interaction gives rise to a ``fifth-force" between CDM particles,
mediated by the DE scalar field, that significantly modifies the
gravitational instability processes through which cosmic structures
develop, both in the linear and in the non-linear regimes \citep[see
  e.g.][]{Farrar2007,Maccio_etal_2004,baldi2010}.

The background evolution of cDE cosmologies is described by the following set of dynamic
equations:
\begin{eqnarray}
\label{klein_gordon}
\ddot{\phi } + 3H\dot{\phi } +\frac{dV}{d\phi } &=&
\sqrt{\frac{2}{3}}\eta _{c}(\phi ) \frac{\rho _{c}}{M_{{\rm Pl}}} \,,
\\
\label{continuity_cdm}
\dot{\rho }_{c} + 3H\rho _{c} &=& -\sqrt{\frac{2}{3}}\eta _{c}(\phi
)\frac{\rho _{c}\dot{\phi }}{M_{{\rm Pl}}} \,, \\
\label{continuity_baryons}
\dot{\rho }_{b} + 3H\rho _{b} &=& 0 \,, \\
\label{continuity_radiation}
\dot{\rho }_{r} + 4H\rho _{r} &=& 0\,, \\
\label{friedmann}
3H^{2} &=& \frac{1}{M_{{\rm Pl}}^{2}}\left( \rho _{r} + \rho _{c} +
\rho _{b} + \rho _{\phi} \right)\,,
\end{eqnarray}
where an overdot represents a derivative with respect to the cosmic
time $t$, $H\equiv \dot{a}/a$ is the Hubble function, $V(\phi)$ is the
scalar field self-interaction potential, $M_{\rm Pl}\equiv
1/\sqrt{8\pi G}$ is the reduced Planck mass, and the subscripts
$b\,,c\,,r$ indicate baryons, CDM and radiation, respectively. The
``coupling function" $\eta_{c}(\phi)$ sets the strength of the
interaction between the DE scalar field and CDM particles, while the
quantity $\dot{\phi} \eta_c(\phi)$ indicates the direction of the
energy-momentum flow between the two fields.
If $\dot{\phi} \eta_c(\phi) > 0$, the transfer
of energy momentum occurs from CDM to DE, while the opposite happens
when $\dot{\phi} \eta_c(\phi) < 0$, due to the convention assumed in
Eqs.~(\ref{klein_gordon}) and (\ref {continuity_cdm}). This implies
that the CDM particle masses, $M_c$, change in time, according to the
following equation:
\begin{equation}
\label{mass}
\frac{d \ln M_{c}}{dt} = -\sqrt{\frac{2}{3}}\eta _{c}(\phi )\dot{\phi} \,.
\end{equation}
Following \citet{Amendola_2004}, the coupling function can be
expressed as:
\begin{equation}
\label{cfunc}
\eta_c(\phi) = \eta_0e^{\eta_1 \phi} \, .
\end{equation}
According to this equation, two different cases are possible: if
$\eta_1=0$, the coupling is constant, i.e. $\eta_c(\phi)=\eta_0$;
otherwise, the interaction strength changes in time due to the
dynamical evolution of the DE scalar field $\phi $
\citep[see][]{baldi2011d}. Both possibilities will be investigated in
the present work.  We will also consider two possible forms for the
scalar field self-interaction potential, $V(\phi)$:
\begin{itemize}
\item a ``runaway" function given by the exponential potential
  \citep{lucchin1985,wetterich1988}:
\begin{equation}
\label{exponential}
V(\phi) = Ae^{-\alpha \phi} \, ,
\end{equation}
\item and a confining function given by the SUGRA potential
  \citep{brax1999}:
\begin{equation}
\label{SUGRA}
V(\phi) = A\phi ^{-\alpha }e^{\phi ^{2}/2} \, ,
\end{equation}  
\end{itemize}
where the field $\phi$ is expressed in units of the reduced Planck
mass. The exponential model has a scalar field that always rolls down
the potential with positive velocity $\dot \phi>0$, reaching the
normalization value, $\phi=0$, at the present time. The SUGRA model is
instead assumed to start with the scalar field at rest in the
potential minimum $\phi _{0} = \sqrt{\alpha}$ in the early Universe,
and the dynamics induced by the coupling allows for an inversion of
the direction of motion of the field during the cosmic expansion
history. This determines a ``bounce" of the DE equation of state,
$w_{\phi}$, on the cosmological constant barrier, $w_{\phi}=-1$, at
relatively recent times, that significantly impacts the number density
of massive CDM haloes at different cosmic epochs. Due to this peculiar
feature, this kind of model has been termed the {\em bouncing} cDE
scenario \citep[see][for a detailed discussion]{baldi2012}.

\begin{figure}
\includegraphics[width=0.45\textwidth]{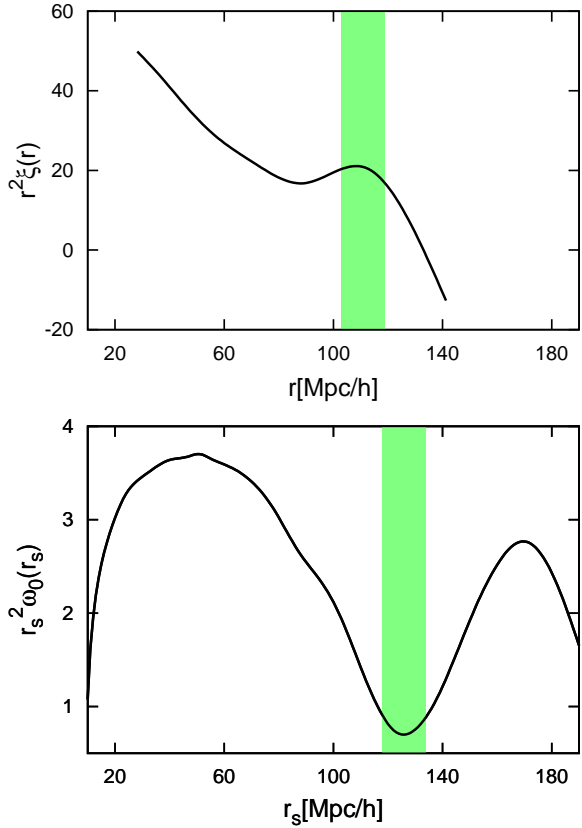}
\caption{The standard two-point correlation function of DM haloes,
  $\xi(r)$ ({\em upper panel}), compared to the band-filtered
  correlation, $\omega_0(r_s)$ ({\em lower panel}). Both statistics
  are computed in the $\Lambda$CDM scenario, at $z=0$. The green
  shaded bands mark the regions where the BAO feature is localized: a
  peak at $\sim110$\Mpch in the $\xi(s)$ correlation function, a dip
  at $\sim130$\Mpch in $\omega_0(r_s)$.}
\label{fig:csi_ns}
\end{figure}
The time evolution of linear density perturbations in the coupled CDM
fluid and in the uncoupled baryonic component ($\delta _{c,b}\equiv
\delta \rho _{c,b}/\rho _{c,b}$) is described at sub-horizon scales
and in Fourier space by the following equations:
\begin{eqnarray}
\label{gf_c}
\ddot{\delta }_{c} &=& -2H\left[ 1 - \eta _{c}\frac{\dot{\phi
  }}{H\sqrt{6}}\right] \dot{\delta }_{c} + 4\pi G \left[ \rho
  _{b}\delta _{b} + \rho _{c}\delta _{c}\Gamma _{c}\right] \,, \\
\label{gf_b}
\ddot{\delta }_{b} &=& - 2H \dot{\delta }_{b} + 4\pi G \left[ \rho
  _{b}\delta _{b} + \rho _{c}\delta _{c}\right]\,,
\end{eqnarray}
where, for simplicity, the dependence of the coupling function $\eta
(\phi)$ has been omitted.  The factor $\Gamma _{c}\equiv 1 + 4\eta
_{c}^{2}(\phi )/3$ in Eq.~(\ref{gf_c}) represents the fifth-force
mediated by the DE scalar field, while the second term in the first
square bracket at the right-hand-side of Eq.~(\ref{gf_c}) is the
extra-friction on CDM fluctuations arising as a consequence of
momentum conservation \citep[see e.g.][]{Amendola_2004, pettorino2008,
  baldi2010, baldi2011a}.  A detailed description of the background
and of the evolution of linear perturbations of all the cosmologies
considered in this work can be found in \citet{CoDECS} and
\citet{baldi2012}.

%%%%%%%%%%%%%%%%%%%%%%%%%%%%%%%%%%%%%%%%%%%%%%%%%%%%%%%%%%%%%%%%%%%%%%%%%%%%%%%

\section{The N-body simulations}
\label{N-body}

Similarly to all the other DE models proposed in the last decades, the
cDE scenarios investigated in this work are barely distinguishable
from the $\Lambda$CDM model from their background expansion history.
Furthermore, despite the fact that such cosmologies determine a
different growth of linear density perturbations as compared to the
concordance $\Lambda $CDM scenario, such effect is highly degenerate
with standard cosmological parameters like $\Omega_{M}$ and, most
importantly, $\sigma_{8}$. Hence, to discriminate between these
cosmologies and the $\Lambda$CDM one, it is useful to exploit also the
non-linear regime of structure formation, which requires to rely on
large numerical simulations to accurately model all the
non-linearities involved.  The analysis described in the following
sections is based on the public halo catalogues of the {\small CoDECS}
simulations \citep{CoDECS}, the largest N-body runs for cDE
cosmologies to date. The range of cosmological models considered in
the {\small CoDECS} project, which includes the $\Lambda$CDM model as
fiducial reference, and the constant, exponential and bouncing cDE
cosmologies, are summarized in Table~\ref{tab:models} along with their
respective parameters.  The viability of these models in terms of CMB
observables has yet to be properly investigated, in particular for
what concerns the impact of variable-coupling and bouncing cDE models
on the large-scale power of CMB anisotropies. Although such analysis
might possibly lead to tighter bounds on the coupling and on the
potential functions than the ones allowed in the present work, here we
are mainly interested in understanding the impact of cDE scenarios on
the statistical properties of large-scale structures at late times,
with a particular focus on the role played by non-linear effects, and
we deliberately choose quite extreme values of the models parameters
in order to maximize the effects under investigation.

The simulation suite considered in this work is the {\small L-CoDECS}
series, a set of collisionless N-body runs in a cosmological box of 1
\Gpch on a side, with $1024^3$ CDM particles and $1024^3$ baryonic
particles, a mass resolution of $m_{c}(z=0) = 5.84\times 10^{10}$
M$_{\odot }/h$ and $m_{b} = 1.17\times 10^{10}$ M$_{\odot }/h$ for CDM
and baryons, respectively, and a force resolution of $\epsilon_{g} =
20$ kpc$/h$.  The simulations have been carried out with a modified
version of the widely used parallel Tree-PM N-body code {\small
  GADGET} \citep{springel2005gadget2, baldi2010}.

The set of cosmological parameters at $z=0$ assumed in the {\small
  CoDECS} project are: $H_0=70.3\rm\,km\,s^{-1}\,Mpc^{-1}$,
$\Omega_{\rm CDM}=0.226$, $\Omega_{\rm DE}=0.729$, $\sigma_{8}=0.809$,
$\Omega_b=0.0451$ and $n_s=0.966$, consistent with the ``CMB-only
Maximum Likelihood" results of WMAP7 \citep[][]{wmap7}. All the
{\small L-CoDECS} simulations are normalized at $z=z_{\rm CMB}$, using
the same initial linear power spectrum, and then rescaling the
resulting displacements to the starting redshift of the simulations,
$z_{i} = 99$, with the specific growth factor $D_{+}(z)$ obtained for
each cosmological model by numerically solving
Eqs.~(\ref{gf_c}, \ref{gf_b}).

The CDM haloes have been identified with a standard Friends-of-Friends
(FoF) algorithm \citep[][]{davis1985}, using a linking length $\lambda
= 0.2\times \bar{d}$, where $\bar{d}$ is the mean interparticle
separation. The results presented in this paper have been obtained
using mass-selected sub-halo catalogues, composed by the
gravitationally bound substructures that the algorithm {\small
  SUBFIND} \citep{springel2001} identifies in each FoF halo. For all
the simulations, we adopted the following mass range: $M_{\rm
  min}<M<M_{\rm max}$, where $M_{\rm min}=2.5\cdot10^{12} M_\odot/h$
and $M_{\rm max}=3.6\cdot10^{15}, 1.1\cdot10^{15}, 4.9\cdot10^{14},
2.6\cdot10^{14}, 1.8\cdot10^{14} M_\odot/h$ at $z=0,0.55,1,1.6,2$,
respectively. In this paper, we only use mock subhalo catalogues and
do not investigate which type of galaxies or clusters are more suited
to achieve the highest accuracy and optimize our results. We defer a
detailed discussion on this topic to a future paper.
\begin{figure}
\includegraphics[width=0.45\textwidth]{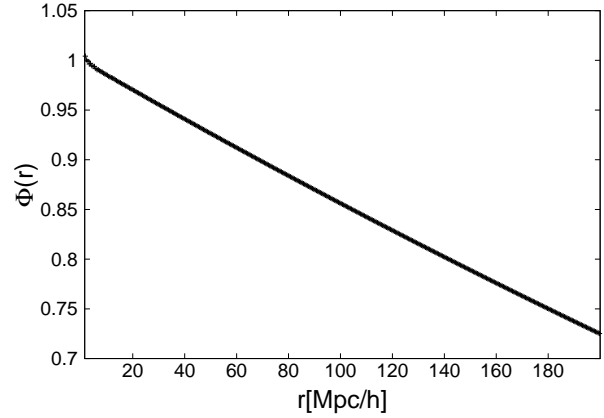}
\caption{The survey geometry function $\Phi(r)$ computed with
  Eq.~(\ref{random}). The bin used for its computation is d$r=0.5$
  \Mpch and $N_R = 5N_D$, where $N_D$ and $N_R$ are the number of
  haloes and random points, respectively. As can be seen, $0 \leq
  \Phi(r) \leq 1$ and $\Phi(r) \rightarrow 1$ when $r\rightarrow 0$.}
\label{fig:sg}
\end{figure}

%%%%%%%%%%%%%%%%%%%%%%%%%%%%%%%%%%%%%%%%%%%%%%%%%%%%%%%%%%%%%%%%%%%%%%%%%%%%%%%
%%%%%%%%%%%%%%%%%%%%%%%%%%%%%%%%%%%%%%%%%%%%%%%%%%%%%%%%%%%%%%%%%%%%%%%%%%%%%%%

\begin{figure*}
\includegraphics[width=\textwidth]{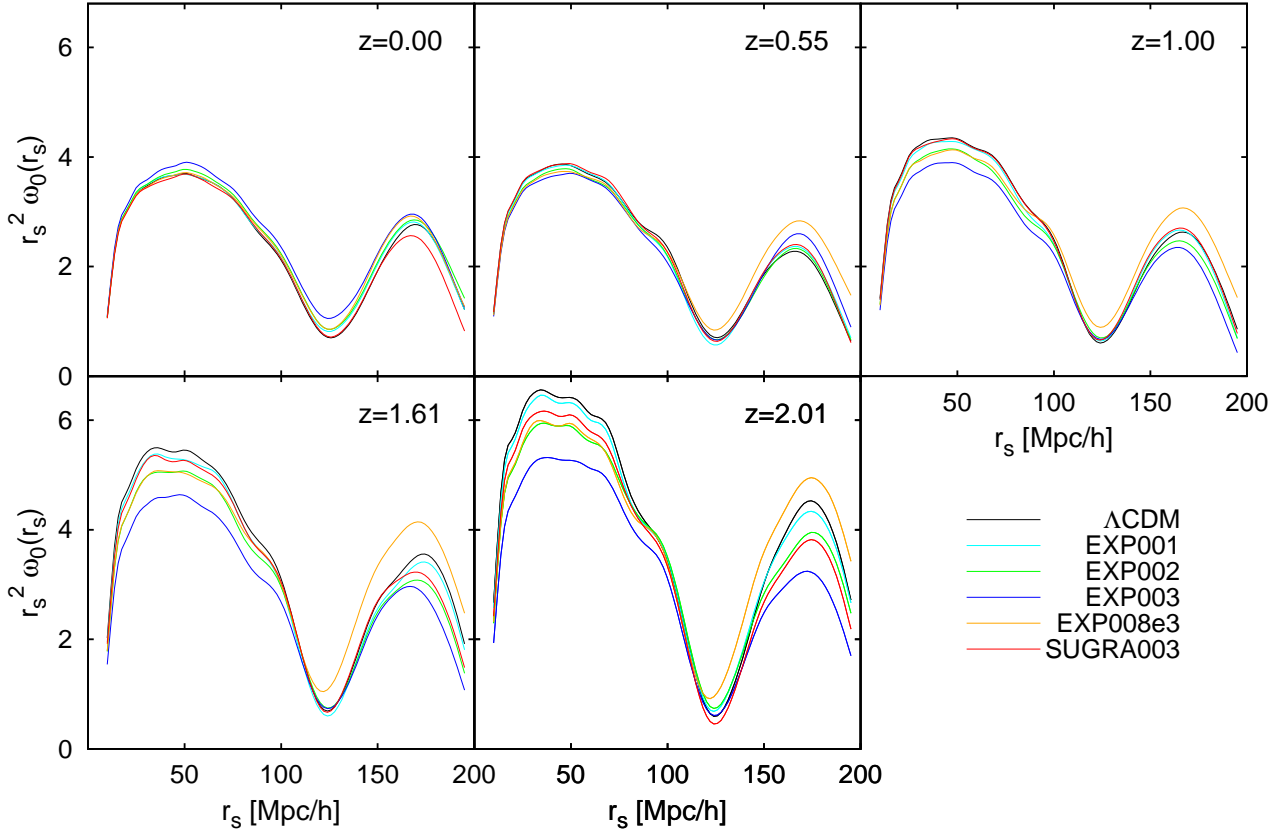}
\caption{The band-filtered correlation function $\omega_0(r_s)$ for
  the different cosmological models considered in the {\small CoDECS}
  simulations, measured in real-space. The BAO signal is clearly seen
  as a single dip. Different panels refer to different redshifts, as
  labelled.}
 \label{fig:ns_all}
\end{figure*}

\section {The band-filtered correlation function}
\label{new_stat_th}
The simplest way to analyze the spatial properties of the LSS of the
Universe is to use two-points statistics. The probability, $dP$, of
finding a pair with one object in the volume $dV_1$ and the other in
the volume $dV_2$, separated by a co-moving distance $r$, is given by:
\begin{equation}
dP=n^2[1+\xi(r)]dV_1dV_2 \, ,
\end{equation}
where $\xi(r)$ is the two-point correlation function and $n$ is the
number density of the objects. \citet{xu2010} introduced a new
statistic, $\omega_l(r_s)$, to analyze the LSS and, in particular, the
BAO, that presents some good advantages with respect to the
traditional estimators, $\xi(r)$, and its Fourier transform,
$P(k)$. In this paper, we will focus on the monopole
$\omega_{l=0}(r_s)$, that is directly linked to the traditional
$\xi(r)$ and dominates the signal with respect to all higher-order
terms \citep[see][for more details]{xu2010}.  The basic idea of this
new statistic is to band-filter the acoustic information in the
correlation function as follows:
\begin{equation}
\label{ns}
\omega_0(r_s) = 4\pi \int_{0}^{r_s} \frac{dr}{r_s} \left (
\frac{r}{r_s} \right )^{2}\xi(r)W(r,r_s) \, .
\end{equation}
The two-parameter filter function, $W(r, r_s)$, must be compact and
compensated ($\int r^2 W(r, r_s) dr =0$) with a characteristic scale
$r_s$ \citep{padmanabhan2007}.  A convenient choice is to use the
following analytical expression:
\begin{equation}
\label{filter}
\left\{
\begin{array}{ll}
  W(x)= 4x^{2} (1-x)\left (\frac{1}{2}-x \right) & 0<x<1 \, ,   \\
  W(x)= 0      & \textrm{otherwise}.
\end{array}
\right.
\end{equation}
where $x=(r/r_s)^3$. Figure \ref{fig:csi_ns} compares the standard
two-point correlation function of DM haloes ({\em upper panel}) with
the band-filtered correlation ({\em lower panel}). Both statistics are
computed in the $\Lambda$CDM model, at $z=0$. While the BAO appears as
a single {\em peak}, at $\sim110$\Mpch, in the traditional $\xi(r)$,
it is instead a single {\em dip}, at $\sim130$\Mpch, in the new
statistic $\omega_0(r_s)$.  It has been shown that $\omega_0(r_s)$
combines the advantages of both the correlation function and the power
spectrum approaches, at the same time \citep{xu2010, blake2011}.
Indeed, $\omega_0(r_s)$ is insensitive to small-scale power, which is
difficult to model due to non-linear effects, and it is also
insensitive to large-scale power, where it is slightly affected by the
integral constraint. It is also easy to measure, as it only requires
to weight the pair counts, without having to bin the data.  If we
adopt the natural estimator for the two-point correlation function:
\begin{equation}
\hat{\xi}(r,\mu)=\frac{DD(r, \mu)}{RR(r, \mu)}-1 \, ,
\label{corr_func}
\end{equation} 
where $DD(r, \mu)$ and $RR(r, \mu)$ are the number of galaxy-galaxy
and random-random pairs with separation $r$, and line-of-sight angle
$\arccos(\mu)$, then from Eqs.~(\ref{filter}) and (\ref{corr_func}) we
can convert the integral given by Eq.~(\ref {ns}) into the following
expression:
\begin{equation}
\label{new_stat}
\omega_0(r_s) =\frac{2}{n_D N_D} \sum_{i=1}^{N} \frac
      {W(r_i/r_s)}{\Phi(r_i, \mu_i)} \, ,
\end{equation}
where $N_D$ and $n_D$ are the number and number density of objects,
respectively, and $\Phi(r, \mu)$ is a normalization function that
encodes the geometry of the survey and depends on the number of data
points.  In general, $0 \leq \Phi(r, \mu) \leq1$ and when $r, \mu
\rightarrow 0$, $\Phi(r, \mu) \rightarrow 1$.  For simulations with
periodic boundary conditions, $\Phi(r, \mu)=1$, as the volume is
effectively infinite. The factor 2 appears because every pair is
counted twice.  The survey geometry function, $\Phi(r, \mu)$, can be
estimated by filling the survey volume with randomly distributed
points, taking into account that their number has to be much larger
than the number of data points, in order to keep the shot noise in the
random pairs smaller than in the object pairs. Hence, the function
$\Phi(r, \mu)$ can be derived counting the random pairs, $RR$, and
using the following equation:
\begin{equation}
RR(r, \mu)=2\pi n_D N_D r^{2}\Phi(r, \mu)\textrm{d}r \textrm{d}\mu \,
,
\label{random}
\end{equation} 
where d$r$ and d$\mu$ are the bins used to count the $RR$ pairs in the
survey. Notice that only the RR pairs, used to estimate the function
$\Phi(r, \mu)$, are measured in finite bins, while the data are not
binned, unlike the traditional correlation function measurements.

%%%%%%%%%%%%%%%%%%%%%%%%%%%%%%%%%%%%%%%%%%%%%%%%%%%%%%%%%%%%%%%%%%%%%%%%%%%%%%%%%%

\section{Results}
\label{results}

\subsection{Measuring the BAO scale in $\omega_0(r_s)$}
\label{model}
\begin{figure}
\includegraphics[width=0.45\textwidth]{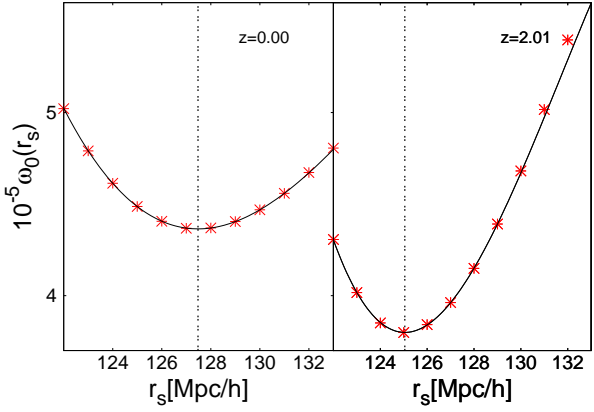}
\caption{The band-filtered correlation function, $\omega_0(r_s)$, in
  the $\Lambda$CDM cosmology, at $z=0$ and $z\sim2$.  Red stars show
  the measured $\omega_0(r_s)$, while the black lines are the best-fit
  polynomial model. The vertical dotted lines mark the position of the
  BAO scale.}
 \label{fig:ns_lcdm}
\end{figure}
To construct mock halo catalogues from the snapshots of the {\small
  CoDECS} simulations, we locate virtual observers at $z=0$ and put
each simulation box at a comoving distance corresponding to its output
redshifts \citep[see][for a detailed discussion]{marulli2012b,
  bianchi2012}. With the above method, each snapshot box represents a
finite volume of the Universe, and the survey geometry function,
$\Phi(r, \mu)$, is not constant, as it would be if we measured the
distances relying on the periodic boundary conditions. However, our
simple mock geometry guarantees that $\Phi(r, \mu)$ is almost
independent of $\mu$, as we have explicilty verified. We use
Eq.~(\ref{random}) to estimate this survey geometry function,
generating random catalogues five times larger than the halo ones and
counting the number of random pairs in bins of d$r=0.5$ \Mpch.  The
result is shown in Fig. \ref{fig:sg}.  We measure the band-filtered
correlation function, $\omega_0(r_s)$, using Eqs.~(\ref{new_stat}) and
(\ref{random}), and linearly interpolating the function $\Phi(r)$ that
we computed as just discussed. Fig. \ref{fig:ns_all} shows the
real-space $\omega_0(r_s)$ of DM haloes, as a function of redshift ($z
= \{0.00,\, 0.55,\, 1.00,\, 1.61,\, 2.01$\}), measured in the range
$10 \leq r_s[$\Mpch$] \leq 190$. The black solid lines show the
$\Lambda$CDM predictions, while the other curves refer to the cDE
models of the {\small CoDECS} simulations, as reported by the labels.
At small scales, $r_s \lesssim 50$ \Mpch, we confirm the results found
by \citet{marulli2012} using the standard two-point correlation
function. More specifically, we find that, at $z=0$, the small-scale
halo spatial properties predicted by the cDE models are very similar
to the $\Lambda$CDM ones. Going to higher redshifts, it is easier to
discriminate between our DE models that predict different shapes of
$\omega_0(r_s)$. Future experiments, such as Euclid, will be able to
accurately constrain the spatial properties of high redshift objects,
thus helping in discriminating between alternative cosmological
scenarios.  

However, in this paper we are more interested in the horizontal shift
of the BAO peak due to cDE non-linear dynamics, as this is not
degenerate with $\sigma_8$, nor with the linear galaxy bias. So we
will focus on low redshifts, where this effect is larger.  To
accurately localize the sound horizon scale in $r_s$-space, with the
band-filtered correlation function (i.e the scale $r_s$ corresponding
to the minimum value of $\omega_0(r_s)$), we fitted the function
$\omega_0(r_s)$ around the dip using a polynomial function. As we are
only interested in the position of the BAO feature, it is not
necessary to model the shape of $\omega_0(r_s)$ at all
scales. Instead, we restrict our analyisis around the BAO dip:
$|r_{s0}-r|\lesssim5$\Mpch where $r_{s0}$ is
the scale corresponding to the minimum value of $\omega_0(r_{s0})$. In
the above range, the shape of band-filtered correlation function can
be accurately modeled with a polynomial function of third order:
\begin{equation}
P_3(r_s)=\sum_{i=0}^{4}a_i.r_s^i \, ,
\label{pol3}
\end{equation}
where $a_i$ are the coefficients of the polynomial. We estimate the
$r_{s0}$ values for all the cosmological models considered in the
{\small CoDECS} simulations and at different redshifts, fitting
$\omega_0(r_s)$ with the model given by Eq.~(\ref{pol3}) and finding
the local minimum of the best-fit polynomials. As a case study,
Fig. \ref{fig:ns_lcdm} shows the band-filtered correlation function
around the dip, measured in the $\Lambda$CDM halo mock catalogues at
$z=0$ and $z\sim2$: red stars are the computed $\omega_0(r_s)$, while
the black lines show the best-fit polynomials. The vertical dotted
lines mark the position of the BAO scale. As clearly shown in
Fig. \ref{fig:ns_lcdm}, the polynomial function is an excellent model
for $\omega_0(r_s)$ around the minimum.
\begin{figure}
\includegraphics[width=0.45\textwidth]{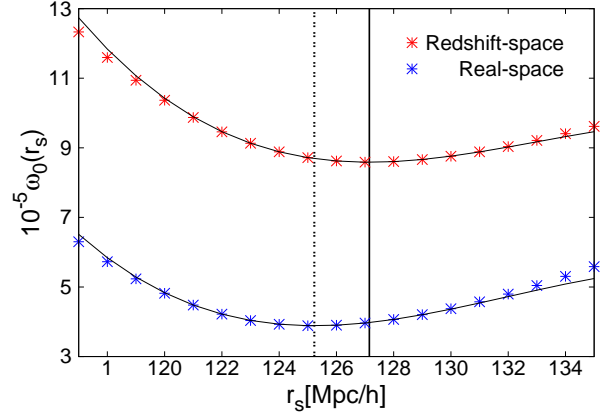}
\caption{The band-filtered correlation functions in real- (blue stars)
  and redshift-space (red stars), for the $\Lambda$CDM cosmological
  model at $z=1$. Black lines are their fitting polynomials. The
  dotted (solid) vertical black line indicates the position of the BAO
  scale in real-space (redshift-space).}
\label{fig:r_z_space}
\end{figure}

%%%%%%%%%%%%%%%%%%%%%%%%%%%%%%%%%%%%%%%%%%

\subsection{The shift of the BAO scale}

\subsubsection{The impact of non-linear dynamics}
\label{non_lin}

The BAO position changes slightly through cosmic time, as can be
directly observed using numerical simulations. This shift is mainly
driven by physical effects, as non-linear structure formation at large
scales, that moves the position of the BAO signal to smaller scales
relative to CMB-calibrated predictions. This occurs as a consequence
of gravitational instability being non local \citep[see
  \eg][]{crocce2008}. Therefore, at higher redshifts, where
non-linearities have a small influence, the BAO peak in the two-point
correlation function is located at larger scales than at lower
redshifts. This effect can also be noticed in the band-filtered
correlation function, with the difference that the BAO feature in
$r_s$-space, i.e $r_{s0}$, is now located at smaller scales for higher
redshifts, due to the effect of the filter, given by
Eq.(\ref{filter}). This can be clearly seen in Fig. \ref{fig:ns_lcdm}:
the position of the BAO scale changes in time, from
$r_{s0}\simeq127.5$ at $z=0$ to $r_{s0}\simeq125$ at $z=2$. In
$r_s$-space, both the scale and the amplitude of the BAO feature
increase as redshifts decrease.
\begin{figure}
\includegraphics[width=0.45\textwidth]{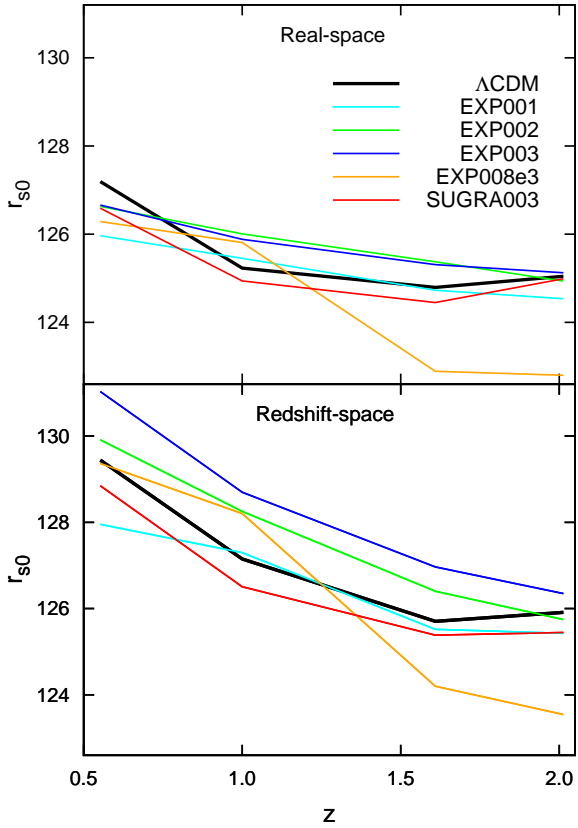}
\caption{Evolution of the BAO scale in the band-filtered correlation
  function, due to non-linearities for real- (upper panel) and
  redshift-space (lower panel) from $z\sim0.5$ to $z=2$.}
\label{fig:rs1}
\end{figure}

%%%%%%%%%%%%%%%%%%%%%%%%%%%%%%%%%%%%%%%%%%%%%%%%%%%%%%%%%%%%%%%%%%%%%%%%%%%%%%%

\subsubsection{The impact of redshift-space distortions}
\label{rsd}

RSD induced by galaxy peculiar motions further shift the position of
the BAO scale \citep{smith2008}. To construct redshift-space halo mock
catalogues, we estimate the observed redshift of each object via the
following equation:
\begin{equation}
z_{obs}=z_c+\frac{\nu_{\Vert}}{c}(1+z_c) \, ,
\label{z_obs}
\end{equation}
where $z_c$ is the cosmological redshift due to the Hubble recession
velocity at the comoving distance of the halo, $\nu_{\Vert}$ is the
line-of-sight component of its centre of mass velocity and $c$ is the
speed of light. Fig. \ref{fig:r_z_space} shows the effect of RSD on
the BAO feature, for the $\Lambda$CDM model at $z=1$. Blue and red
stars show the real- and redshift-space band-filtered correlation
functions, respectively. The vertical dotted and solid black lines
indicate the position of the BAO scale in real- and redshift-space,
respectively. The amplitude enhancement of $\omega_0(r_s)$ in
redshift-space is caused by the Kaiser effect and is directly
proportional to the linear growth rate of cosmic structures
\citep{marulli2012}.  Here, we are more interested in the horizontal
shift, as this is not degenerate with $\sigma_8$, allowing to break
the degeneracy between the effects arising as a consequence of the
DE-CDM interactions and standard cosmological parameters.  As shown in
Fig. \ref{fig:r_z_space}, the BAO scale is shifted by $\sim 2$ \Mpch
from real- to redshift-space.
\begin{figure}
\includegraphics[width=0.45\textwidth]{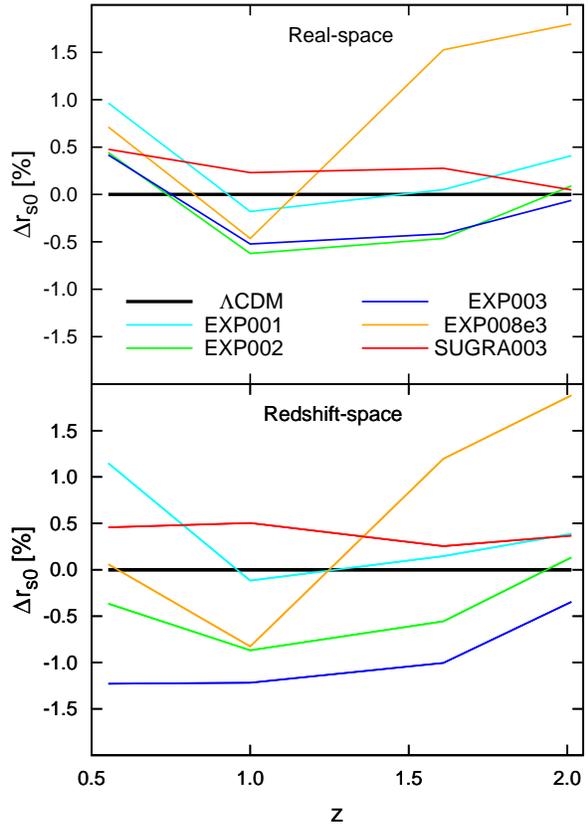}
\caption{The percentage difference between cDE and $\Lambda$CDM model
  predictions of the evolution of the BAO scale in the band-filtered
  correlation function, due to non-linearities for real- and
  redshift-space from $z\sim0.5$ to $z=2$. }
\label{fig:rs1_error}
\end{figure}

%%%%%%%%%%%%%%%%%%%%%%%%%%%%%%%%%%%%%%%%%%%%%%%%%%%%%%%%%%%%%%%%%%%%%%%%%%%%%%%

\subsection{The BAO scale as a dynamical probe}

The shift of the BAO scale due to non-linearities and RSD depends
directly on the underlying cosmology. When the BAO feature is used as
a geometric standard ruler, it is necessary to accurately correct for
the shift, to avoid systematics when constraining cosmological
parameters. Most importantly, the correction should be different for
each cosmological model to test. On the other hand, if the sound
horizon scale is measured with sufficient accuracy, the BAO shift can
provide a dynamical probe to discriminate between different
cosmological models.

The interaction between DE and CDM shifts the matter-radiation
equality to higher redshifts, as CDM dilutes faster relative to the
$\Lambda$CDM case and there was more CDM in the past \citep[see
  \eg][]{amendola2012b}. Moreover, in cDE models the evolution of
linear and non-linear density perturbations is significantly altered,
as compared to the standard $\Lambda$CDM scenario. Specifically, as
discussed in \S \ref{cDE}, the DE coupling induces a time evolution of
the mass of CDM particles as well as a modification of the growth of
structures, determined by the combined effect of a long-range
fifth-force, mediated by the DE scalar degree of freedom, and of an
extra-friction, arising as a consequence of momentum conservation. At
large scales, these effects change the location and the amplitude of
the BAO scale. Therefore, the clustering properties of CDM haloes at
the sound horizon scale, in particular the combination of the BAO
shifts due to non-linearities and RSD, might show specific signatures,
allowing to distinguish a cDE Universe from the $\Lambda$CDM scenario.

\subsubsection{The cDE BAO in real- and redshift-space}
\label{rs0_cde}

Using our mock halo catalogues, we can quantify the effect of
non-linear dynamics in the band-filtered correlation function, for the
different cDE models of the {\small CoDECS} project.
Fig. \ref{fig:rs1} shows the evolution of the position of the BAO
scale, $\mathrm{r_{s0}}$, in real- (upper panel) and redshift-space
(lower panel), as a function of redshift and cosmological model. In
both panels we can see that $\mathrm{r_{s0}}$ increases, due to
non-linearities, as redshift decreases.  Comparing the upper and lower
panel, we can see the impact of the RSD on the position of
$\mathrm{r_{s0}}$ for every cosmological model from $z\sim0$ to
$z\sim2$, i.e. the shifts of $\mathrm{r_{s0}}$ to higher scales from
real- to redshift-space as described in \S \ref{rsd}.

In Fig. \ref{fig:rs1_error}, we show the percentage difference between cDE and
$\Lambda$CDM predictions, $\mathrm{\Delta
  r_{s0}=100\cdot(r_{s0,cDE}-r_{s0,\Lambda CDM})/r_{s0,\Lambda CDM}}$,
where $\mathrm{r_{s0,\Lambda CDM}}$ is the $\Lambda$CDM BAO scale, and
$\mathrm{r_{s0,cDE}}$ its counterpart for every cDE model.
The percentage differences between the values of
$\mathrm{r_{s0}}$ in $\Lambda$CDM and cDE models are quite small,
$\lesssim2\%$ in the whole redshift range. The EXP008e3 model predicts
the largest difference, that rises from $\sim1\%$ to $\sim2\%$ at
$z\sim1.5$ and $z\sim2$, respectively. The $\mathrm{r_{s0}}$ difference
in the SUGRA003 scenario is always positive ($\lesssim0.5\%$) and
almost constant up to $z\sim0.5$.

\begin{figure}
\includegraphics[width=0.45\textwidth]{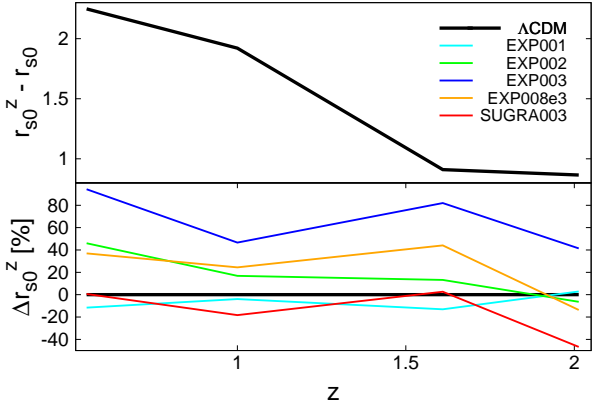}
\caption{{\em Upper panel}: The evolution of the difference between the sound
horizon scale in redshift- and real-space for the $\Lambda$CDM scenario. 
{\em Bottom panel}: Percentage difference between cDE models and 
$\Lambda$CDM model predictions obtained from the Eq.~(\ref{diff_rs0}).}
\label{fig:rs2}
\end{figure}

The most interesting result of Fig. \ref{fig:rs1} is the amplification
effect of RSD. While, in real-space, $\mathrm{\Delta r_{s0}}$ is small
for all the cDE models considered (except for EXP008e3, at high
redshifts), the predictions of the different cDE models are much more
clearly distinguishable in redshift-space. To better quantify this
effect, in Fig. \ref{fig:rs2} we show the difference between the sound
horizon scale in redshift- and real-space. RSD move the sound horizon
at larger scales (as highlighted by the black line in the upper
panel, that shows the $\Lambda$CDM prediction), an effect that
monotonically increases at low redshifts. More interestingly, the
effect of RSD is strongly model-dependent, as can be clearly seen in
the lower panel of Fig. \ref{fig:rs2}, that shows the percentage difference
between $\mathrm{\Delta r_{s0}}$ in $\Lambda$CDM and cDE models: 
\begin{equation}
\mathrm{\Delta
  r^z_{s0}=100\cdot\frac{(r^z_{s0,cDE}-r^{}_{s0,cDE})-(r^z_{s0,\Lambda
      CDM}-r^{}_{s0,\Lambda CDM})}{{(r^z_{s0,\Lambda
        CDM}-r^{}_{s0,\Lambda CDM})}}} \, ,
\label{diff_rs0}
\end{equation}
where $\mathrm{r^z_{s0,\Lambda CDM}}$ and $\mathrm{r^{}_{s0,\Lambda
    CDM}}$ are the positions of the BAO feature in redshift- and
real-space respectively, in the $\Lambda$CDM scenario, while
$\mathrm{r^z_{s0,cDE}}$ and $\mathrm{r^{}_{s0,cDE}}$ are their
respective counterparts for every cDE scenario.

For constant coupling models, the RSD effect is directly proportional
to the coupling strength, $\beta$: the larger is $\beta$, the most the
BAO scale is shifted to large scales. Indeed, the strongest effect is
found for the EXP003 model, over the whole redshift range, for which
the BAO shift due to RSD is up to $\sim90\%$. On the other hand, the
impact of RSD is almost negligible for the EXP001 and SUGRA003
models. We note that $\mathrm{\Delta r_{s0}}$ can be negative for
these models, but the effect is small considering the statistical
uncertainties.

%%%%%%%%%%%%%%%%%%%%%%%%%%%%%%%%%%%%%%%%%%%%%%%%%%%%%%%%%%%%%%%%%%%%%%%%%%%%%%%

\subsubsection{The impact of redshift errors}

As we have seen in the previous sections, non-linear small-scale
effects can ``propagate'' to larger scales, significantly shifting the
sound horizon in a cosmology-dependent way. Hence, it is mandatory to
investigate how redshift errors can impact the BAO feature. Here we
consider only Gaussian redshift errors, typical of both spectroscopic
and photometric galaxy surveys, and do not investigate the impact of
the so-called catastrophic errors (the ones caused by the
misidentification of one or more spectral features). Indeed, pure
catastrophic outliers with a flat distribution simply reduce the
amplitude of $\omega_0(r_s)$ at all scales, not affecting the position
of $r_{s0}$. On the other hand, BAO shifts possibly induced by {\em
  systematic} misidentifications of spectral features need to be
estimated with mock galaxy catalogues, taking into account the
specific observational effects of the survey under investigation.

\begin{figure}
\includegraphics[width=0.45\textwidth]{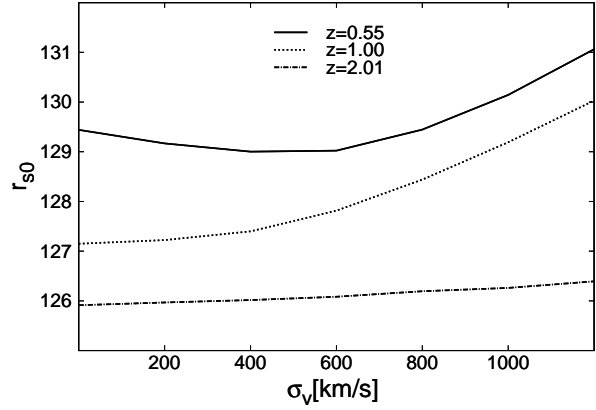}
\caption{The impact of redshift errors on the position of the BAO
  scale, for the $\Lambda$CDM model at different reshifts ($z=0.55$ -
  solid line, $z=1$ - dotted line, $z=2$ - dot-dashed line).}
\label{fig:sigmaV}
\end{figure}

We estimate the redshift of each mock halo with a modified version of Eq.~\ref{z_obs}:
\begin{equation}
z_{obs}=z_c+\frac{\nu_{\Vert}}{c}(1+z_c)+\frac{\sigma_v}{c}  \, ,
\label{sigma_v}
\end{equation}
where the new term $\sigma_v$ is the random error in the measured
redshift (expressed in km/s). To assess the impact of redshift errors
on the position of $\mathrm{r_{s0}}$, we measure the band-filtered
correlation in halo mock catalogues with the following Gaussian
redshift errors: $\sigma_v = \{0, 200, 400, 600, 800, 1200\}$ km/s. In
Fig \ref{fig:sigmaV}, we show our results only for the $\Lambda$CDM
scenario, as the effect of redshift errors does not depend on the
cosmological model.  The impact of Gaussian errors results to be
negligible for $\sigma_v\lesssim 600$ km/s and, at high redshifts
($z\gtrsim1.5$), even for larger values of $\sigma_v$. On the other
hand, for $z\lesssim1$ and $\sigma_v \gtrsim 600$ km/s, the sound
horizon systematically shifts to larger scales as redshift errors
increase. Typical spectroscopic galaxy surveys have quite accurate
redshift measurements ($\sigma_v<100$ km/s), hence the systematic
effect of redshift errors at the BAO scale is tiny in these
cases. However, larger redshift errors, typical of photometric
surveys, could introduce a non-negligible shift in the sound horizon
scale, that should be accurately corrected when extracting
cosmological constraints from the BAO feature.  Interestingly, if we
consider the required redshift error limits of the Euclid surveys,
i.e. $\sigma_z \le 0.001(1 + z)$, which corresponds to 600 km/s at $z
= 1$ \citep{laureijs2011}, the error on the location of
$\mathrm{r_{s0}}$ is $\sim0.5$\%, i.e. small enough to distinguish cDE
models from the $\Lambda$CDM scenario.

%%%%%%%%%%%%%%%%%%%%%%%%%%%%%%%%%%%%%%%%%%%%%%%%%%%%%%%%%%%%%%%%%%%%%%%%%%%%%%%

\section{Conclusions} 
\label{concl}

In this paper, we quantified the impact of DE interactions on the
clustering of CDM haloes at large scales, using the band-filtered
correlation function $\omega_0(r_s)$. We used the public halo
catalogues from the {\small CoDECS} simulations, a suite of large
N-body runs for cDE models. We accurately estimated the sound horizon
scale in our mock halo catalogues, fitting the function
$\omega_0(r_s)$ with third order  polynomials around the BAO scale.
We investigated the effects of non-linear dynamics both in real- and
redshift-space, in the ranges $10\lesssim r_s \lesssim 190$ and $0\leq
z\leq2$. The main results of our analysis are the following.

\begin{itemize}

\item We detected BAO shifts in $\omega_0(r_s)$, relative to the
  linear predictions, for $\Lambda$CDM and cDE cosmologies: non-linear
  effects shift the sound horizon to larger scales, an effect that
  monotonically increases moving to lower redshifts.

\item RSD significantly amplify the non-linear BAO shift, at all
  redshifts. On the other hand, Gaussian redshift errors $\lesssim 600$
  km/s have a negligible impact at the BAO scale.

\item The combined BAO shift due to non-linearities and RSD is
  strongly model dependent. Thus, if the BAO scale is accurately
  measured, it can provide a tool to distinguish between $\Lambda$CDM
  and cDE models.

\end{itemize}

To conclude, besides the common use as a standard ruler, the BAO
feature can be exploited also as a dynamical probe to disentangle
between cosmological scenarios characterized by different non-linear
dynamics. The BAO shift due to cDE dynamics is generally small, but
RSD significantly amplify the effect, expecially at low
redshifts. Moreover, the shift of the BAO scale is not degenerate with
$\sigma_8$ and, consequently, with the other paratemeters degenerate
in turn with $\sigma_8$, like the linear galaxy bias and the total
mass of neutrinos \citep{lavacca2009, kristiansen2010,
  marulli2011}. Thus, the BAO feature provides a strong and unbiased
probe for detecting new dark interactions, and can be of great help in
removing the $\sigma_8$-degeneracy of other intependent observables,
like the halo mass function \citep{cui2012} and the clustering
anisotropies \citep{marulli2012}.

\section*{acknowledgments}
We warmly thank C. Carbone for helpful discussions and suggestions. We
acknowledge the support from grants ASI-INAF I/023/12/0, PRIN MIUR
2010-2011``The dark Universe and the cosmic evolution of baryons: from
current surveys to Euclid". VDVC is supported by the Erasmus Mundus
External Cooperation Window Lot 18. MB is supported by the Marie Curie
Intra European Fellowship ``SIDUN" within the 7th European Community
Framework Programme and also acknowledges support by the DFG Cluster
of Excellence ``Origin and Structure of the Universe'' and by the
TRR33 Transregio Collaborative Research Network on the
``DarkUniverse''.

\bibliographystyle{mn2e} \bibliography{bib,baldi_bibliography}

\end{document}